\documentclass[aps,prl,twocolumn,a4paper,english]{revtex4}
\usepackage{xcolor}
\usepackage{amsmath,epsfig,amssymb,subfigure,bm,dsfont}
\usepackage{lipsum} 
\usepackage{blindtext}
\usepackage{graphicx}
\usepackage{titlesec}

\begin{document}
\title{Collective resonance of $D$ states in rubidium atoms probed by optical two-dimensional coherent spectroscopy}
\author{Danfu Liang}
\affiliation{Department of Physics, Florida International University, Miami, FL 33199}

\author{Yifu Zhu}
\affiliation{Department of Physics, Florida International University, Miami, FL 33199}

\author{Hebin Li\thanks{2}}
\email{hebin.li@fiu.edu}
\affiliation{Department of Physics, Florida International University, Miami, FL 33199}

\begin{abstract}
Collective resonance of interacting particles has important implications in many-body quantum systems and their applications. Strong interactions can lead to a blockade that prohibits the excitation of a collective resonance of two or more nearby atoms. However, a collective resonance can be excited with the presence of weak interaction and has been observed for atoms in the first excited state ($P$ states). Here, we report the observation of collective resonance of rubidium atoms in a higher excited state ($D$ states) in addition to the first excited state. The collective resonance is excited by a double-quantum four-pulse excitation sequence. The resulting double-quantum two-dimensional (2D) spectrum displays well-isolated peaks that can be attributed to collective resonances of atoms in $P$ and $D$ states. The experimental one-quantum and double-quantum 2D spectra can be reproduced by a simulation based on the perturbative solutions to the optical Bloch equations, confirming collective resonances as the origin of the measured spectra. The experimental technique provides a new approach for preparing and probing collective resonances of atoms in highly excited states. 
\end{abstract}

\maketitle

Collective resonance of multiple particles has important implications for quantum information science \cite{Bernien2017,Mazurenko2017}, quantum metrology \cite{Ludlow2008,Swallows2011}, strongly correlated systems \cite{Trotzky2012}, photosynthesis \cite{Engel2007}, and fundamental studies of many-body physics \cite{Weiner1989}. A well-known example is Dicke state \cite{Dicke1954} which is a coherent collection of $N$ atoms. A hallmark phenomenon of Dicke states is superradiance, in which the collective resonance of $N$ atoms produces a cooperative spontaneous emission whose intensity scales with $N^2$ instead of $N$. In the case of Dicke states, the formation of collective resonance can be mediated by interacting with a common optical field and does not necessarily require interatomic interactions. On the other hand, collective effects due to strong interatomic interactions have been extensively studied in ensembles of highly excited Rydberg atoms, promising quantum applications such as nonclassical light sources \cite{Peyronel2012,Maxwell2013,Dudin2012} and quantum gates \cite{Jaksch2000, Lukin2001}. Owing to their large dipole moments, highly excited Rydberg atoms interact strongly and lead to the blockade of excitations for surrounding atoms in the blockade regime. This phenomenon, known as the Rydberg blockade, is the key for many applications of Rydberg atoms. However, the blockade also prohibits the simultaneous excitation of two interacting Rydberg atoms to form a collective resonance \cite{Gaetan2009,Urban2009} which is necessary to manipulate entanglement states of two or more atoms. The excitation of collective resonance is possible with the presence of weak interatomic interaction and has been observed for atoms in the first excited state ($P$ states) \cite{Dai2012,Gao:16,Lomsadze2018,Yu2019a,Yu2019,PhysRevA.92.053412,Bruder2019,doi:10.1063/5.0052982}. It is of interest to induce and probe collective resonances of atoms in higher excited states. 

Optical two-dimensional coherent spectroscopy (2DCS), which is an optical analog of two-dimensional nuclear magnetic resonance \cite{Ernstbook}, has been demonstrated as a powerful tool to study many-body correlations and interactions in various systems \cite{Cundiff2013,Li2017}. In particular, double-quantum 2DCS was used to probe collective resonances due to weak dipole-dipole interactions in potassium (K) and rubidium (Rb) atomic vapors \cite{Dai2012,Gao:16,Lomsadze2018,Yu2019a}. In double-quantum 2DCS, the excitation pulses create a double-quantum coherence between the ground state and the doubly-excited state that can be a collective state of two atoms. However, the signals from all excitation pathways cancel out if the two atoms do not interact. The presence of interaction breaks the symmetry so that the cancellation is incomplete, resulting in a nonzero double-quantum signal \cite{Dai2012,Gao:16}. The double-quantum 2DCS provides sensitive detection of the collective resonances induced by weak interatomic interactions. The technique can also be extended to detect multi-quantum coherence associated with collective resonances of multiple atoms \cite{PhysRevA.92.053412, Yu2019, doi:10.1063/5.0052982}. The observed collective resonances are collective states of two or more atoms in the $P$ states but not higher excited states. 

In this letter, we report the observation of collective resonances of Rb atoms in the $D$ state in addition to the $P$ state in an Rb atomic vapor. The collective resonances are created and detected by a four-pulse double-quantum excitation sequence in an optical 2DCS experiment. The excitation pulses generate double-quantum coherences between an initial state and a doubly-excited state. For atoms initially prepared in the $P$ state, the collective resonances of two atoms in the $D$ state can be generated and the resulting double-quantum signals are unambiguously manifested in the 2D spectrum as an isolated peak. Both the one- and double-quantum spectra involving the $D$ state are presented. The experimental spectra can be reproduced by simulation based on the perturbative solutions to the optical Bloch equations, confirming collective resonances as the origin of the observed signals. This work provides a new experimental approach for generating and manipulating collective resonances of atoms in highly excited states, including Rydberg states, for potential applications of quantum many-body systems. 

\begin{figure}
\includegraphics[width=\columnwidth]{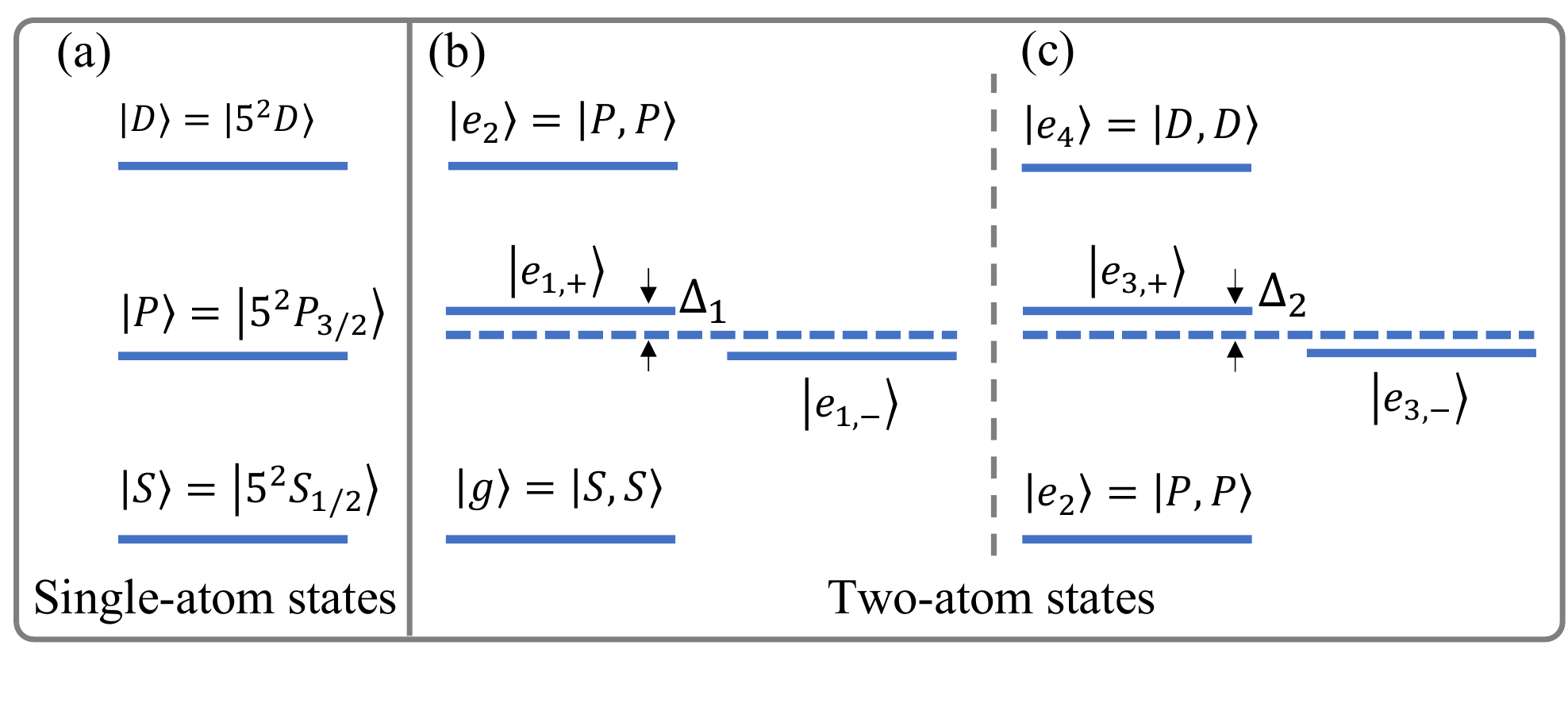}
\caption{(a) Relevant single-atom energy levels of Rb atoms, including ground state $|S\rangle$, singly excited state $|P\rangle$, and doubly excited state $|D\rangle$. (b) Collective states of two Rb atoms are $|g\rangle=|S,S\rangle$ as the initial state, $|e_{1,\pm}\rangle=\frac{1}{\sqrt{2}}(|PS\rangle\pm|PS\rangle)$, and $|e_2\rangle=|P,P\rangle$. (c) Collective states of two Rb atoms are $|e_2\rangle=|P,P\rangle$ as the initial state, $|e_{3,\pm}\rangle=\frac{1}{\sqrt{2}}(|DP\rangle\pm|PD\rangle)$, and $|e_4\rangle=|D,D\rangle$.
}\label{fig:1}
\end{figure}

The collective resonances of Rb atoms are measured in our experiment. The relevant Rb energy levels are $|S\rangle=|5^2S_{1/2}\rangle$, $|P\rangle=|5^2P_{3/2}\rangle$, and $|D\rangle=|5^2D\rangle$, as shown in Fig. \ref{fig:1}(a). The $|5^2P_{1/2}\rangle$ state is outside the laser bandwidth and the hyperfine levels are not resolved in our measurements. For single atoms, the excitation pulses with a central wavelength of 778 nm can generate a double-quantum coherence between $|S\rangle$ and $|D\rangle$, which leads to two off-diagonal peaks in the double-quantum 2D spectrum \cite{Gao:16,Yu2019a}. When two atoms are considered in their joint Hilbert space, the collective states of $|S\rangle$ and $|P\rangle$ form a four-level system, as shown in Fig. \ref{fig:1}(b), including the ground state $|g\rangle=|S,S\rangle$, singly excited states $|e_{1,\pm}\rangle=\frac{1}{\sqrt{2}}(|P,S\rangle\pm|S,P\rangle)$, where state $|e_{1,-}\rangle=\frac{1}{\sqrt{2}}(|P,S\rangle-|S,P\rangle)$ is a dark state that cannot be excited, and a doubly excited state $|e_2\rangle=|P,P\rangle$. Therefore the system can be considered as a three-level ladder system with energy shift $\Delta_1$ for singly excited state. For two atoms initially in the ground state $|g\rangle$, the double-quantum excitation can generate a double-quantum coherence between $|g\rangle$ and $|e_2\rangle$ and give rise to a diagonal peak in the double-quantum 2D spectrum \cite{Gao:16,Yu2019a}, providing evidence for dipole-dipole interaction induced collective resonances of Rb atoms in the $P$ state. Here, we further consider two atoms initially in the doubly excited state $|e_2\rangle$. As shown in Fig. \ref{fig:1}(c), they can similarly be excited into states $|e_{3,+}\rangle=\frac{1}{\sqrt{2}}(|D,P\rangle+|P,D\rangle)$ by a single-quantum excitation and state $|e_4\rangle=|D,D\rangle$ by a double-quantum excitation. The double-quantum coherence between $|e_2\rangle$ and $|e_4\rangle$ can result in an isolated peak in the double-quantum 2D spectrum as the evidence for the collective resonance of two atoms in the $D$ states.  

\begin{figure}[htp]
\centering
\includegraphics[width=\columnwidth]{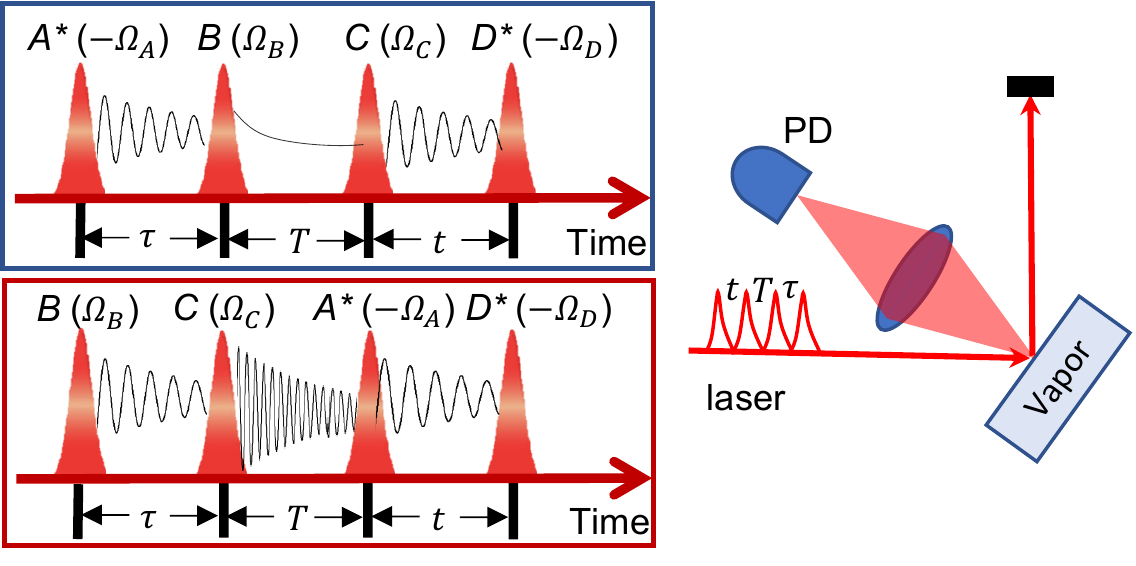}
\caption {Experimental schematic of optical 2DCS. Four copropagating pulses are incident on the window of a vapor cell and the fluorescence signal is detected by a photodetector (PD). The upper and lower insets show the one-quantum and double-quantum excitation pulse sequences, respectively.}\label{fig:2}
\end{figure}

Optical 2DCS experiment is implemented in a collinear setup based on acousto-optic modulators (AOMs) \cite{Nardin2013,Yu2019,Yu2019a}. As shown in Fig. \ref{fig:2}, four co-propagating excitation pulses are incident on the window of an Rb atomic vapor cell. The cell is heated to $170$ $^{\circ}$C in an oven and the atomic density is $2.06 \times 10^{14}$ cm$^{-3}$. The excitation pulses are derived from the output of a Ti:sapphire femtosecond oscillator by using a nested Mach-Zehnder interferometer \cite{Nardin2013}. The pulses are about 200 fs in duration at a repetition rate of 78 MHz. The spectrum has a central wavelength of 778 nm and a bandwidth of 6 nm at the full width half maximum (FWHM). The total power of four pulse trains at the cell window is 45 mW. The combined beam is focused by a lens to a spot on the window with a size of $\sim100$ $\mu$m in diameter (FWHM). The pulses are labeled $A^*$, $B$, $C$, and $D^*$ and the time delays are $\tau$, $T$, and $t$ for the first, second, and third delays, respectively, between the pulses. Each pulse is phase modulated by an AOM at a slightly different frequency $\Omega_A=80.107$ MHz, $\Omega_B=80.104$ MHz, $\Omega_C=80.0173$ MHz and $\Omega_D=80$ MHz respectively. Pulses $A^*$ and $D^*$ are considered conjugated in our excitation schemes so their corresponding modulation frequencies are $-\Omega_A$ and $-\Omega_D$. Two excitation pulse sequences are used. As shown in Fig. \ref{fig:2}, the upper inset shows the pulse sequence for one-quantum excitation and the lower inset for double-quantum excitation. Both excitation sequences generate a fourth-order nonlinear signal. For instance, in the double-quantum excitation, the first pulse, $B$, generates a single-quantum coherence between the ground and singly-excited states; the second pulse, $C$, converts the single-quantum coherence to a double-quantum coherence between the ground and doubly-excited states; the third pulse, $A^*$, converts the double-quantum coherence to a third-order single-quantum coherence; and the fourth pulse, $D^*$, turns the single-quantum coherence into a fourth-order population which emits a fluorescence signal. The signal is detected by a photodetector (PD) and demodulated by a lock-in amplifier. The fourth-order nonlinear signal is selected by the lock-in amplifier at the reference frequency $\Omega_S=\Omega_B-\Omega_A+\Omega_C-\Omega_D=14.3$ kHz. The signal is recorded as a function of two time delays and 2D Fourier-transformed into the frequency domain to generate a 2D spectrum. A one-quantum 2D spectrum is generated by scanning $\tau$ and $t$ in the one-quantum excitation sequence. A double-quantum 2D spectrum requires to scan $T$ and $t$ in the double-quantum excitation sequence.

\begin{figure}
\includegraphics[width=\columnwidth]{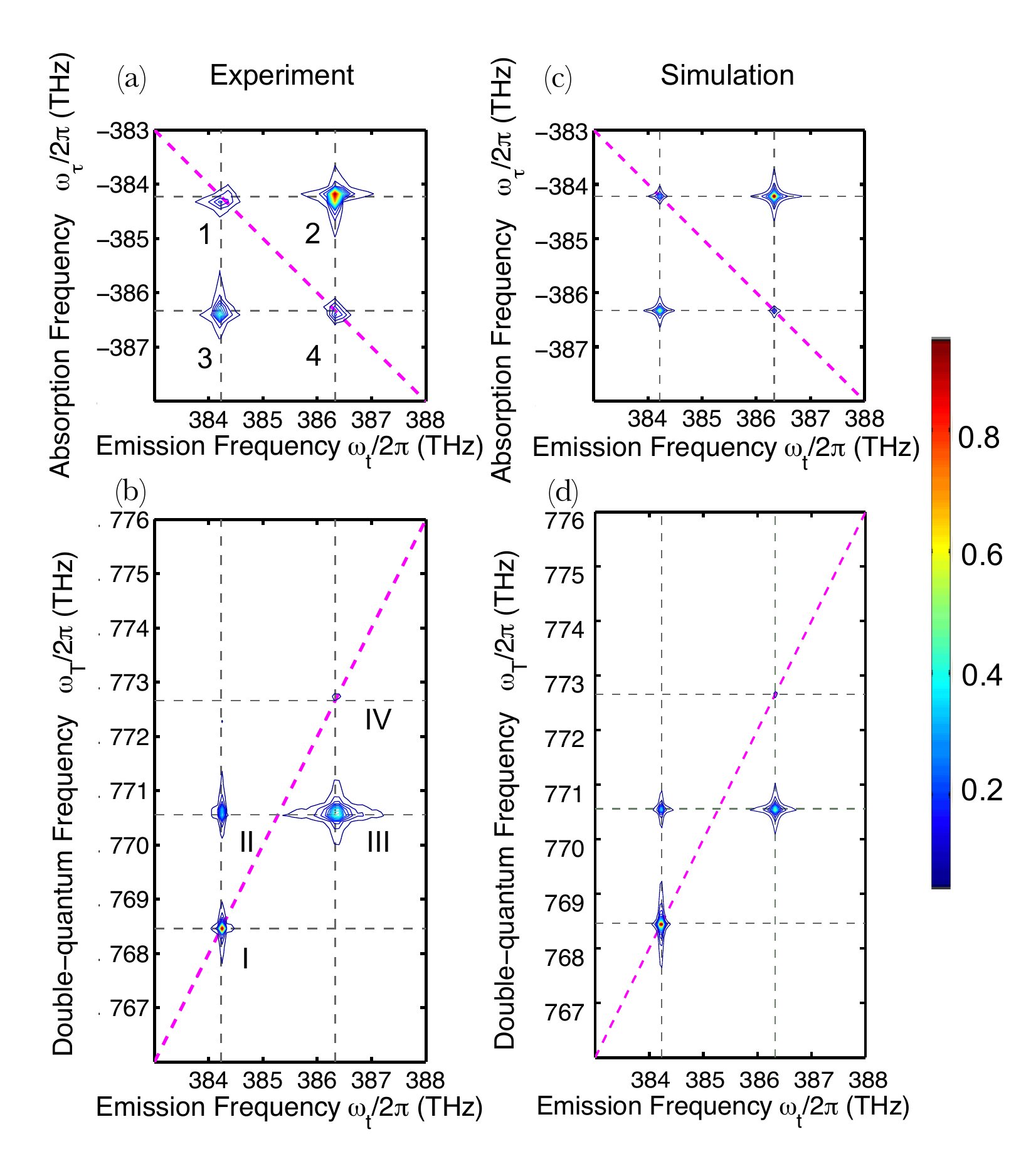} 
\caption{Experimental (a) one-quantum and (b) double-quantum 2D spectra and simulated (c) one-quantum and (d) double-quantum 2D spectra. In all spectra, the amplitude is plotted with the maximum normalized to one.}\label{fig:3}
\end{figure}
 
The acquired one-quantum and double-quantum 2D spectra are shown in Fig. \ref{fig:3}(a) and \ref{fig:3}(b), respectively. The spectral amplitude is plotted with the maximum normalized to one. All time delays are scanned for 10 ps corresponding to a frequency resolution of 0.1 THz so the hyperfine levels and two $5^2D$ states ($J=1/2, 3/2$) are not resolved in the measurement. The one-quantum spectrum was obtained with the excitation pulse sequence shown in the upper inset in Fig. \ref{fig:2}. The absorption frequency $\omega_\tau$ axis and the emission frequency $\omega_t$ axis correspond to the time delays $\tau$ and $t$, respectively. The diagonal peak 1 is due to the transition from $|S\rangle$ to $|P\rangle$, while peak 4 corresponds to the transition from $|P\rangle$ to $|D\rangle$. There are also two off-diagonal peaks 2 and 3 due to the coupling between peaks 1 and 4. The existence of peak 4 suggests that the first pulse can excite some atoms into the $|P\rangle$ state. Therefore, the subsequent optical 2DCS experiment needs to account for both $|S\rangle$ and $|P\rangle$ states as possible initial states. For atoms that are initially prepared in the $|P\rangle$ state, it is then possible to excite collective resonances of atoms in the $|D\rangle$ state by using the double-quantum excitation pulse sequence shown in the lower inset Fig. \ref{fig:2}. The resulting double-quantum 2D spectrum, as shown in Fig. \ref{fig:3}(b), has a double-quantum frequency $\omega_T$ axis and an emission frequency $\omega_t$ axis corresponding to the time delays $T$ and $t$, respectively. Peaks I, II, and III have been previously reported in double-quantum 2D spectra of Rb atoms \cite{Gao:16,Yu2019a}. They are attributed to the double-quantum signal from the excitation of atoms initially in the $|S\rangle$ state. Peaks II and III are associated with the single-atom state $|D\rangle$ while peak I is contributed by the two-atom collective state $|e_2\rangle$. In this experiment, we observed an additional peak labeled as IV. This peak has a double-quantum frequency that is twice the transition frequency from $|P\rangle$ to $|D\rangle$. The double-quantum signal associated with peak IV is attributed to the excitation of atoms initially in the $|P\rangle$ state. A high excitation density is required to prepare a sufficient number of atoms in the $|P\rangle$ state. Peak IV was absent in the previously reported double-quantum 2D spectra due to the relatively low laser power used in the experiment.   

\begin{figure}
\includegraphics[width=\columnwidth]{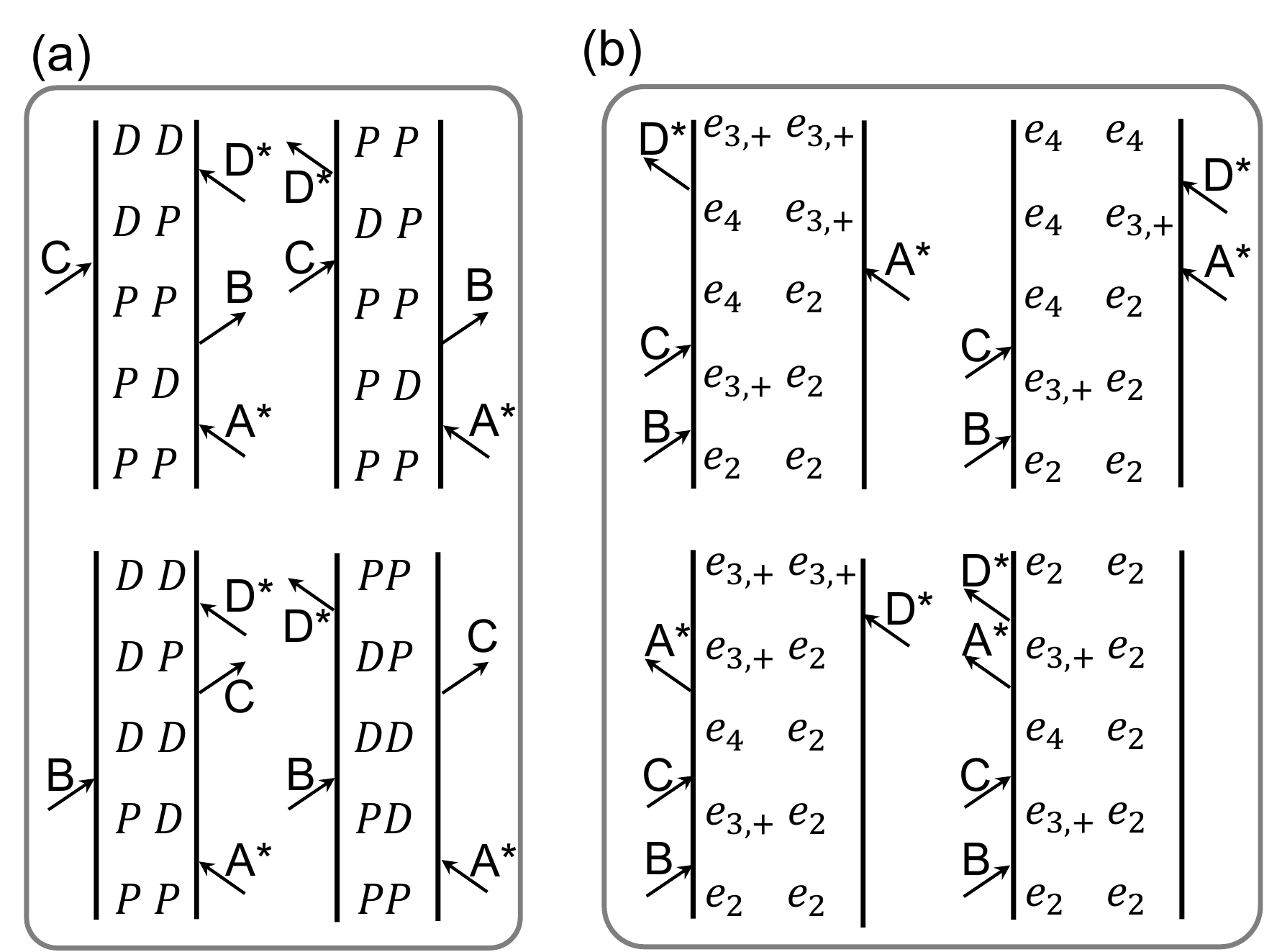} 
\caption{Double-sided Feynman diagrams showing excitation pathways that contribute to (a) peak 4 in one-quantum 2D spectrum and (b) peak IV in double-quantum 2D spectrum.}\label{fig:4}
\end{figure}

Both one-quantum and double-quantum 2D spectra can be reproduced by a simulation based on the perturbative solutions to the optical Bloch equation. Under the excitation of the pulse sequences in Fig. \ref{fig:2}, each peak in the spectra is contributed by several excitation pathways that can be represented by double-sided Feynman diagrams. The pathways in Fig. \ref{fig:4}(a) and \ref{fig:4}(b) contribute to peak 4 in the one-quantum 2D spectrum and peak IV in the double-quantum 2D spectrum, respectively. The pathways associated with other peaks are shown in Supplemental Material (SM) \cite{supplement}. Each excitation pathway generates a fourth-order population which can be calculated from the double-sided Feynman diagram. The pathways in Fig. \ref{fig:4}(a) lead to a fourth-order population in state $|P\rangle$ that emits one photon as well as state $|D\rangle$ that emits two photons as a fluorescence signal. The signal in peak 4 is the sum of the contributions from all pathways in Fig. \ref{fig:4}(a) and can be calculated as \cite{supplement}
\begin{eqnarray}
S_{\text{4}}(\omega_\tau,\omega_t)=\frac{2S_0\rho^{(0)}_{PP}}{(\omega_{\tau}-\omega_{PD}+i\Gamma_{DP})(\omega_{t}-\omega_{DP}+i\Gamma_{DP})},
\end{eqnarray}
where $S_0=-\frac{E_AE_BE_CE_D\mu_{PD}^4}{16\hbar}$. Here $E_{A,B,C,D}$ are the electric field amplitudes, $\hbar$ is the reduced Planck constant, $\rho^{(0)}_{PP}$ is the initial population in state $|P\rangle$, $\mu_{ij}$ is dipole moment, $\Gamma_{ij}$ is the relaxation rate, and $\omega_{ij} = \omega_i-\omega_j$ is the frequency difference between states $|i\rangle$ and $|j\rangle$. Similarly, based on the pathways shown in Fig. \ref{fig:4}(b), the signal in peak IV can be calculated as
\begin{eqnarray}
S_{\text{IV}}(\omega_T,\omega_t)=\frac{3S_0\rho^{(0)}_{e_2e_2}}{\omega_T-\omega_{e_4e_2}+i\Gamma_{e_4e_2}} \nonumber \\ (\frac{1}{\omega_t-\omega_{e_{3,+}e_2}+i\Gamma_{e_{3,+}e_2}}    -\frac{1}{\omega_t-\omega_{e_4e_{3,+}}+i\Gamma_{e_4e_{3,+}}}), \label{eq:peakIV}
\end{eqnarray}
where $\rho^{(0)}_{e_2e_2}$ is the initial population in state $|e_2\rangle$. The signals for all other peaks can be calculated based on the doubled-sided Feynman diagrams for the contributing pathways as shown in SM \cite{supplement}. Simulated one-quantum and double-quantum 2D spectra, as shown in Fig. \ref{fig:3}(c) and \ref{fig:3}(d) respectively, are generated from the calculated signals as shown in SM \cite{supplement}. The simulation shows that peak 4 in the one-quantum 2D spectrum is due to the transition from $|P\rangle$ to $|D\rangle$, indicating there is an initial population in state $|P\rangle$ for the 2DCS measurement. Within the initial population in $|P\rangle$, some of the atoms are in the correlated two-atom state $|e_2\rangle$. The double-quantum excitation pulse sequence in Fig. \ref{fig:2} can then access states $|e_{3,+}\rangle$, and $|e_4\rangle$ and generate double-quantum coherence between states $|e_4\rangle$ and $|e_2\rangle$, as illustrated by doubled-sided Feynman diagrams in Fig. \ref{fig:4}(b). The resulting double-quantum signal calculated from Eq. \ref{eq:peakIV} would be zero if the interaction between the two atoms is absent in which case we have $\omega_{e_4e_{3,+}}=\omega_{e_{3,+}e_2}$ and $\Gamma_{e_4e_{3,+}}=\Gamma_{e_{3,+}e_2}$. The existence of peak IV in the double-quantum 2D spectrum is a result of the two-atom states $|e_{3,+}\rangle$, $|e_4\rangle$ and the interaction between the two atoms. 

In summary, we observed collective resonances of Rb atoms in the $D$ and $P$ states in an atomic vapor by using optical 2DCS experiments. Both one-quantum and double-quantum 2D spectra were measured. The one-quantum 2D spectrum shows that some atoms are initially prepared in the $P$ state. The double-quantum 2D spectrum includes signals due to double-quantum coherences between the two-atom collective states $|e_4\rangle$ (two atoms in the $D$ state), $|e_2\rangle$ (two atoms in the $P$ state), and $|g\rangle$ (two atoms in the $S$ state). The double-quantum signal also indicates the interaction between two atoms. The simulated 2D spectra based on the perturbative solutions to the optical Bloch equations agree with the experimental spectra and confirm collective resonances as the origin of the observed double-quantum signals. The developed technique can provide a new approach to prepare and probe collective resonances of atoms in highly excited states including Rydberg states for quantum applications requiring many-body systems.

This material is based upon work supported by the National Science Foundation under Grant No. PHY 1707364.

\bibliography{RbDState}

\end{document}


\title{Supplemental Material: \\Collective resonance of $D$ states in rubidium atoms probed by optical two-dimensional coherent spectroscopy}
\author{Danfu Liang}
\affiliation{Department of Physics, Florida International University, Miami, FL 33199}

\author{Yifu Zhu}
\affiliation{Department of Physics, Florida International University, Miami, FL 33199}

\author{Hebin Li\thanks{2}}
\email{hebin.li@fiu.edu}
\affiliation{Department of Physics, Florida International University, Miami, FL 33199}
\maketitle


In this Supplemental Material, we describe the calculation that reproduces the experimental 2D spectra. The calculation is based on the fourth-order perturbative solutions of the optical Bloch equation. The contributing pathways are represented by the double-sided Feynman diagrams shown in Fig. S1. 

\begin{figure*}[htp]
\centering
\includegraphics[width=0.74\textwidth]{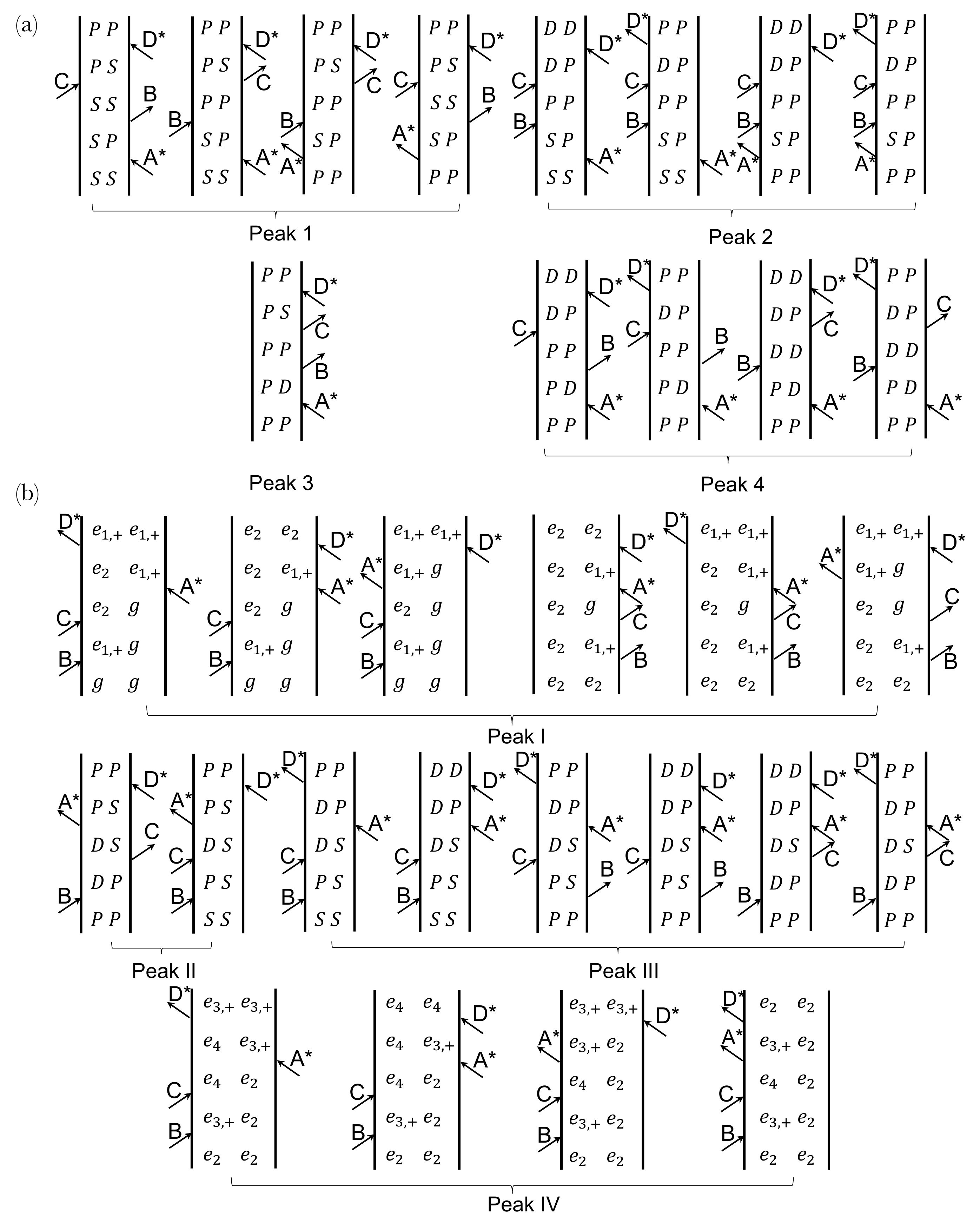}
\caption{Double-sided Feynman diagrams representing (a) excitation pathways contributing each peak in the one-quantum 2D spectrum and (b) excitation pathways contributing each peak in the double-quantum 2D spectrum. The pathways are grouped according to each individual peak. }\label{figS1}
\end{figure*}

The laser excitation of atoms in our experiment can be modeled based on the density matrix formalism and the equation of motion is given by the optical Bloch equation as \cite{Scully1997}
\begin{equation}
    \Dot{\rho}=-\frac{i}{\hbar}[H,\rho]-\frac{1}{2}\{ \Gamma, \rho \}, 
\end{equation}
where $[H,\rho]=H\rho-\rho H$ and $\{\Gamma, \rho\}=\Gamma\rho+\rho\Gamma$ with $H$, $\Gamma$, and $\rho$ being the Hamiltonian, relaxation, and density matrix, respectively. The matrix elements of $H$ are $H_{ij}=\hbar \omega_i \delta_{ij}-\mu_{ij}\mathcal{E}(t)$, where $\hbar\omega_i$ is the energy of state $|i\rangle$, $\mathcal{E}(t)$ is the electric field, $\mu_{ij} (i\neq j)$ is the dipole moment of the transition between states $|i\rangle$ and $|j\rangle$, and $\delta_{ij}$ is the Kronecker delta function. The relaxation matrix elements are $\Gamma_{ij}=\frac{1}{2}(\gamma_i+\gamma_j)$, where $\gamma_i$ and $\gamma_j$ are the population decay rates. To include the decoherence due to pure dephasing processes in addition to population decay, the relaxation matrix elements can be written as $\Gamma_{ij}=\frac{1}{2}(\gamma_i+\gamma_j)+\gamma_{ij}^{ph}$, where $\gamma_{ij}^{ph} (i\neq j)$ is the pure coherence dephasing rate. The equation of motion can be modified as 
\begin{equation}
    \dot{\rho}_{ij}=-\frac{i}{\hbar}\sum_k (H_{ik}\rho_{kj}-\rho_{ik}H_{kj})-\Gamma_{ij}\rho_{ij}. \label{eq:OBE}
\end{equation}
The laser pulses are considered $\delta$-pulses and the electric field of each laser pulse is 
\begin{equation}
    \mathcal{E}_j(t)=E_j \delta (t-t_j) e^{-i(\omega_it+\Omega_i t)}+ c.c.,
\end{equation}
where $\delta (t-t_j)$ is the Dirac delta function, $E_j$ is the electric field amplitude, $t_j$ is the pulse arrival time, and $\Omega_i$ is the phase modulation frequency by AOM for each pulse with $j=A$, $B$, $C$, and $D$. Under the excitation of four pulses $A$, $B$, $C$, and $D$, the resulting fourth-order nonlinear signal is $S^{(4)} \propto \chi^{(4)} \mathcal{E}_A\mathcal{E}_B\mathcal{E}_C\mathcal{E}_D$ with $\chi^{(4)}$ being the fourth-order susceptibility. Considering each pulse is phase modulated by AOM at frequency $\Omega_i$, in general the fourth-order signals due to the excitation of all four pulses have phase modulation frequencies of all possible combination $\pm \Omega_A \pm \Omega_B \pm \Omega_C \pm \Omega_D$. In our experiment, only the signal with the modulation frequency $-\Omega_A + \Omega_B + \Omega_C - \Omega_D$ is detected by the lock-in amplifier. Therefore, we only calculate the signals due to the excitation pathways that result in this particular phase modulation frequency. Pulses $A$ and $D$ are considered conjugated in these pathways. 

Equation (\ref{eq:OBE}) is solved perturbatively to find the fourth-order nonlinear signal. The process can assisted by using double-sided Feynman diagrams \cite{Boyd2003,Shen2002}. Each diagram represents a particular excitation pathway contributing to a term in the perturbation solution. The contributing pathways to each peak are shown in Fig. \ref{figS1}(a) for the one-quantum 2D spectrum and Fig. \ref{figS1}(b) for the double-quantum 2D spectrum. The contributions to the fourth-order nonlinear signal can be calculated for each diagram and summarized to calculate the overall signal. As an example, the fourth-order population $\rho^{(4)}_{PP}$ resulted from the first pathway in Fig. \ref{figS1}(a) can be calculated as
\begin{eqnarray}
\rho_{PP}^{(4)}&=&\frac{-i\mu_{PS}}{2\hbar}e^{-i\Omega_D}\int_{-\infty}^{t_e}\dif t''''e^{-\Gamma_{PP}(t_e-t'''')}E_{D}^{*}\delta(t''''-t_4)e^{i\omega t''''} \nonumber \\
&&\cdot\frac{-i\mu_{SP}}{2\hbar}e^{i\Omega_C}\int_{-\infty}^{t''''}\dif t'''e^{-i(\omega_{SP}-i\Gamma_{SP})(t''''-t''')}E_C\delta(t'''-t_3)e^{i\omega t'''} \nonumber \\
&&\cdot\frac{-i\mu_{SP}}{2\hbar}e^{i\Omega_B}\int_{-\infty}^{t'''}\dif t''e^{-i(\omega_{SS}-i\Gamma_{SS})(t'''-t'')}E_B\delta(t'''-t_2)e^{i\omega t''} \nonumber \\
&&\cdot\frac{-i\mu_{SP}}{2\hbar}e^{-i\Omega_A}\int_{-\infty}^{t''}\dif t'e^{-i(\omega_{SP}-i\Gamma_{SP})(t''-t')}E_{A}^{*}\delta(t'-t_1)e^{i\omega t'}\cdot \rho_{SS}^{(0)}\nonumber \\
&=&\frac{-\mu_{SP}^4}{16\hbar^4}e^{i(-\Omega_A+\Omega_B+\Omega_C-\Omega_D)}E^*_AE_BE_CE^*_D\rho_{SS}^{(0)}\Theta(\tau)\Theta(T)\Theta(t)\Theta(t_e)\nonumber \\
&&\cdot e^{\Gamma_{PP}t_e}e^{-i(\omega_{SP}-\omega)t-\Gamma_{SP}t}e^{-i(\omega_{SS}-\omega)T-\Gamma_{SS}T}e^{-i(\omega_{SP}-\omega)\tau-\Gamma_{SP}\tau}, \label{eq:timesolution}
\end{eqnarray}
where $t_e$ is the fluorescence emission time, $\Theta$'s are the Heaviside step functions, the $\rho_{PP}^{(0)}$ is the initial population of $P$ state, and $\hbar$ is the reduced Planck constant. The pulses are considered as $\delta$-pulses and the rotating wave approximation is
used in the calculation. The fluorescence signal detected by a slow photo-detector is proportional to the population at $t_e=0$. Time delays $\tau$ and $t$ are scanned while the delay $T$ is fixed at  $T=0$.
Under these conditions, the frequency domain signal due to the first pathway can be obtained by Fourier transforming Eq. (\ref{eq:timesolution}) and written in the lab frame as
\begin{equation}
S(\omega_\tau, \omega_t)=\frac{-E_AE_BE_CE_D\rho_{SS}^{(0)}\mu_{PS}^4}{16\hbar^4}
\times\frac{1}{\omega_\tau-\omega_{SP}+i\Gamma_{SP}} \times\frac{1}{\omega_t-\omega_{PS}+i\Gamma_{PS}}.
\end{equation}
The signals from other pathways can be calculated similarly and summarized to obtain the signal for each peak in the one-quantum 2D spectrum, 
\begin{eqnarray}
S_{1}(\omega_\tau, \omega_t)&=&2(S_0-S_1)\mu_{SP}^4 
\frac{1}{\omega_\tau-\omega_{SP}+i\Gamma_{SP}}\times\frac{1}{\omega_t-\omega_{PS}+i\Gamma_{PS}},  \\
S_{2}(\omega_\tau, \omega_t)&=&(S_1-S_0+S_2)\mu_{SP}^2\mu_{PD}^2
\frac{1}{\omega_\tau-\omega_{SP}+i\Gamma_{SP}}\times\frac{1}{\omega_t-\omega_{DP}+i\Gamma_{DP}},  \\
S_{3}(\omega_\tau, \omega_t)&=&(-S_1-S_2)\mu_{SP}^2 \mu_{PD}^2
\frac{1}{\omega_\tau-\omega_{PD}+i\Gamma_{PD}}\times\frac{1}{\omega_t-\omega_{DP}+i\Gamma_{DP}}, \\
S_{4}(\omega_\tau, \omega_t)&=&2S_1 \mu_{PD}^4
\frac{1}{\omega_\tau-\omega_{PD}+i\Gamma_{PD}}\times\frac{1}{\omega_t-\omega_{DP}+i\Gamma_{DP}}, 
\end{eqnarray}
where $S_0=-E_AE_BE_CE_D\rho_{SS}^{(0)}/16\hbar^4$ and $S_1=-E_AE_BE_CE_D\rho_{PP}^{(0)}/16\hbar^4$. Additionally, term $S_2$ is introduced to include the signals that are dropped in the double-sided Feynman diagrams due to the rotating wave approximation. This term appears in the full calculation and should be included when the initial state is the singly excited state $|P\rangle$. 

The signals in the double-quantum 2D spectrum can be calculated from the pathways in Fig. \ref{figS1}(b). Peaks II and III are due to the doubly-excited states of single atoms as well as two-atom states, while peaks I and IV are resulted from the two-atom states. Accounting for the contributions from all excitation pathways, the signals for each peak in the double-quantum 2D spectrum can be calculated as
\begin{eqnarray}
    S_{I}(\omega_T, \omega_t)&=&2(S_0+S_1)\mu_{SP}^4 
    \frac{1}{\omega_T-\omega_{e_2g}+i\Gamma_{e_2g}}\times (\frac{1}{\omega_t-\omega_{e_{1,+}g}+i\Gamma_{e_{1,+}g}}-\frac{1}{\omega_t-\omega_{e_2e_{1,+}}+i\Gamma_{e_2e_{1,+}}}), \\
S_{II}(\omega_T, \omega_t)&=&2(S_0-S_1)\mu_{SP}^2\mu_{PD}^2
\frac{1}{\omega_T-\omega_{DS}+i\Gamma_{DS}}\times\frac{1}{\omega_t-\omega_{PS}+i\Gamma_{PS}},\\ 
S_{III}(\omega_T, \omega_t)&=&2(-S_0+2S_1)\mu_{SP}^2 \mu_{PD}^2
\frac{1}{\omega_T-\omega_{DS}+i\Gamma_{DS}}\times\frac{1}{\omega_t-\omega_{DP}+i\Gamma_{DP}},\\
S_{IV}(\omega_T, \omega_t)&=&3S_1\mu_{PD}^4\frac{1}{\omega_T-\omega_{e_4e_2}+i\Gamma_{e_4e_2}}\times(\frac{1}{\omega_t-\omega_{e_{3,+}e_2}+i\Gamma_{e_{3,+}e_2}} -\frac{1}{\omega_t-\omega_{e_4e_{3,+}}+i\Gamma_{e_4e_{3,+}}}).
\end{eqnarray}

The simulated one-quantum spectrum, as shown in Fig. 3(c), and the double quantum spectrum, as shown in Fig. 3(d), can be generated from Equations (S6 $\sim$ S9) and Equations (S10 $\sim$ S13), respectively. For the energy level scheme in the simulation, the energy shifts are $\Delta_1=\omega_{e_{1,+}g}-\omega_{PS}$ and $\Delta_2=\omega_{e_{3,+}e_2}-\omega_{DP}$. The parameters for the simulated spectra are listed in Table. S1.
\begin{center}
\begin{table}[htb]
    \label{tab:1}
    \begin{tabular}{|c|c|c|c|c|c|c|c|c|c|c|c|c|c|c|c}
    \hline
 Para.& $S_0:S_1:S_2$  &
$\omega_{SP}/2\pi$&$\omega_{DP}/2\pi$&$\Gamma_{DP}/2\pi$ & $\Gamma_{PS}/2\pi$& $\Gamma_{e_{1,+}g}/2\pi$&$\Gamma_{e_2e_{1,+}}/2\pi$&$\Gamma_{e_4e_{3,+}}/2\pi$&$\Gamma_{e_{3,+}e_2}/2\pi$&
$\Delta_1/2\pi$ &$\Delta_2/2\pi$  \\
   \hline
   Value & $3:8:1$  & $384.23$ THz & $386.33$ THz & $30$ GHz&  $10$ GHz & $29$ GHz   &$30$ GHz& $29$ GHz& $18$ GHz & $10$ GHz & $20$ GHz\\
\hline
\end{tabular}
\caption{\label{tab:1} Parameters for the simulation of one-quantum and double-quantum 2D spectra shown in Figs. 3(c) and 3(d).}
\end{table}
\end{center}





\bibliography{supplement}